\documentclass[10pt,twocolumn,amsmath,amssymb,aps,prb]{revtex4-1}
\usepackage[french]{babel}
\usepackage[utf8]{inputenc}
\usepackage[T1]{fontenc}
\usepackage{amsmath}
\usepackage{amssymb}
\usepackage{xcolor}
\usepackage{graphicx}
\usepackage{subfigure}
\usepackage{epstopdf}
\usepackage{hyperref}
\hypersetup{
  colorlinks=true,
  linkcolor=blue,
  citecolor=red,
  linktoc=all,
  pdfauthor=B. Mussard,
  pdftitle=Random Phase Approximation in Projected Oscillator Orbitals,
  pdfdisplaydoctitle=true,
  pdfpagelayout=SinglePage,
  pdfstartview=Fit,
  pdfstartpage=1,
  bookmarksopen=false
}

\newcommand{\braket}[2]{\ensuremath{\langle  #1 \vert #2  \rangle}}
\newcommand{\expect}[3]{\ensuremath{\langle  #1 \vert #2 \vert #3 \rangle}}
\newcommand{\bra}[1]{\ensuremath{\langle #1 \vert}}
\newcommand{\ket}[1]{\ensuremath{\vert #1  \rangle}}
\renewcommand\b[1]{\ensuremath{\mathbf{#1}}}
\newcommand\B[1]  {\ensuremath{\pmb #1}}

\usepackage{tabularx}
\begin{document}
\renewcommand{\arraystretch}{1.2}

\title{Random Phase Approximation in Projected Oscillator Orbitals.}
\author{Bastien Mussard}
\email{bastien.mussard@colorado.edu}
\affiliation{Department of Chemistry and Biochemistry, University of Colorado Boulder, Boulder, CO 80302, USA}

\begin{abstract}
    The projected oscillator orbitals (pOOs) are localized virtual orbitals
    constructed by multiplying localized occupied orbitals by harmonics.
    Following a recent paper by Mussard and \'{A}ngy\'{a}n\cite{us},
    further developments of projected oscillator orbitals are shown,
    notably the equations for pOOs of general order as well as their overlaps are derived.
    The performance of these localized virtual orbitals is demonstrated up to third order.
    It is found that a good fraction of the aug-cc-pVQZ RPA correlation energy
    is recovered by use of a smaller number of pOOs.
    This is especially true where considering only the long-range correlation energy,
    which is important for the description of London dispersion forces. 
\end{abstract}

\maketitle

\section*{Introduction}

In the context of density functional approximations, the functionals
from the so-called fifth rung of the Jacob's ladder not only use
the density, its gradient, and a set of occupied orbitals
but require also \textit{a priori} the knowledge of the set of virtual orbitals.
Such methods have the obvious drawback that the size of the virtual orbital space can be very large.

A solution often used to keep the size of matrices in reasonable limits
employs an auxiliary basis set to expand the occupied-virtual products.
Such approaches are known in quantum chemistry as resolution of identity \cite{Eichkorn:95}
or density-fitting \cite{Whitten:73,Roeggen:08} methods
and similar advantages can be found by use of Cholesky decomposition \cite{Beebe:77} techniques.

Further gain can be achieved by local correlation methods,
which take advantage of the locality of purposefully made orbitals
to reduce the number of significant terms to calculate in a given method.
Hence, both the localization of orbitals\cite{Hoyvik:12,Aqui,Yang2011a,Brani,Zhang,Subot2005,Hess2016,Maynau}
and the derivation of local correlation methods\cite{Yang2011,Yang2012,Christ,Subot2006,Kallay,Hess2017}
have lately been an active area of research.

In the present work it is shown that accurate random phase approximation (RPA) energies can be evaluated
with approximations which avoid any explicit reference to virtual orbitals
and involve quantities that are computable from occupied orbitals alone,
much like what was shown by Surj\'{a}n for MP2\cite{Surjan:05}.
The strategy followed here uses localized virtual orbitals called
the projected oscillator orbitals (pOOs)\cite{us},
in the spirit of the projected atomic orbitals techniques \cite{Pulay:83b,Pulay:86,Boughton:93}.

In a fairly recent paper\cite{us}, we showed that the pOOs,
an original idea by Foster and Boys\cite{Foster:60a,Foster:60b,Boys:66},
could be used in the context of the RPA equations.
The core idea behind the pOOs is to construct a set of virtual orbitals
directly from a set of occupied localized orbitals (LMOs)
by multiplying them with solid spherical harmonics.
The orthogonality of these oscillator orbitals with the occupied space
is ensured by projection to obtain the projected oscillator orbitals.
The pOOs are non-orthogonal among each other which is a source of some complications,
although only to the same extent and in the same manner as in the case of the projected atomic orbitals,
and the work that has been done over the years
in the context of local correlation methods using projected atomic orbitals\cite{Knowles:00} can be used.
As mentioned in another paper\cite{us}, the pOOs show similarities
with other works\cite{Kirkwood:32,Pople:57,Karplus:63a,Karplus:63b,Rivail:78,Rivail:79,Sadlej:71a}.

The main interest of this paper is not to reproduce the full correlation energy with high numerical precision,
but only a well-defined part of it, namely the long-range dynamical correlation energy
which is usually responsible for the London dispersion forces.
It is indeed now well-documented that most of the conventional density functional calculations
in the Kohn-Sham framework are unable to grasp the physics of these long-range forces,
unless special corrections are added
to the total energy\cite{Johnson:05,Johnson:09,Grimme:10,Grimme:11c,Tkatchenko:09,DiStasioJr:14, Reilly:15a}. 
On the other hand it has been demonstrated in earlier works\cite{Toulouse:09,Zhu:10,Angyan:11,Toulouse:11}
that the essential physical ingredients
of London dispersion forces are contained in the range-separated hybrid RPA method,
where the short-range correlation effects are described within a density functional approximation
and the long-range exchange and correlation are handled at the long-range Hartree-Fock and long-range RPA levels.
The range-separated hybrid method has the additional advantage of showing
a more favorable convergance with basis set size\cite{basisconv}.
Note that the use of localized orbitals for dispersion energy calculations has already been proposed
since the early works on local correlation methods \cite{Kapuy:91,Saebo:93,Hetzer:98,Usvyat:07,Chermak:12}.

In the following, new derivations of the projected oscillator orbitals framework are presented
as well as a very brief recall of the local formulation of the random phase approximation equations
before showing the performance on full-range and long-range correlation energies.
Detailed derivations are provided in the Appendix when needed.

\section*{Projected Oscillator Orbitals}

A set of oscillator orbitals (OOs) is produced by
multiplying a set of localized molecular orbitals (LMOs) $\{\ket{i}\}$ by solid spherical harmonics
centered on the barycenter $\b{D}^i$ of the LMOs
(here for the first order solid spherical harmonic):
\begin{align}
\ket{\text{OO}}=(\hat{r}_\alpha-D^i_\alpha)\ket{i}
\end{align}
where $\hat{r}_\alpha$ is the $\alpha=x,y,z$ component of the position operator.
The set of localized virtual orbitals $\{\ket{i_\alpha}\}$ that are the projected oscillator orbitals (pOOs)
is constructed by
projecting the oscillator orbitals out of the occupied space:
\begin{align}
  \label{eq:pOO}
  \ket{\text{pOO}}=\hat{P}\,(\hat{r}_\alpha-D^i_\alpha)\ket{i}
  \doteq\ket{i_\alpha}
\end{align}
where $\hat{P}=(\hat{1}-\sum \ket{k}\bra{k})$
and where the composite index ``$i_\alpha$'' refers to
a pOO generated from the $i$-th LMO by using the $\hat{r}_\alpha$ harmonic.
The pOOs are not orthogonal among each other, and have an overlap (here again between the first order pOOs):
\begin{align}
    S_{i_\alpha j_\beta}
    =\expect{i}{\hat{r}_\alpha\hat{P}\,\hat{r}_\beta}{j}
   &=\expect{i}{\hat{r}_\alpha       \hat{r}_\beta}{j}
\nonumber\\&\quad
-\sum_k^{N_\text{LMO}} \expect{i}{\hat{r}_\alpha}{k}\expect{k}{\hat{r}_\beta}{j}
\end{align}
where $N_\text{LMO}$ is the number of (localized) occupied orbitals.

In Eq.~(\ref{eq:pOO}), the tail
emerging from the projection term $\sum\ket{k}\bra{k}\hat{r}_\alpha\ket{i}$
obviously damages the locality of the pOOs.
The Foster-Boys localization criterion for the LMOs\cite{Foster:60a,Boys:66}
precisely ensures that the sum of the off-diagonal elements of the $\hat{r}_\alpha$ operators
taken between the occupied orbitals is minimized\cite{Resta:06b}.
Hence the pOOs are optimally localized when used in conjunction with
LMOs obtained by use of the Foster-Boys criterion.

The virtual pOOs have by design a nodal surface intersecting the occupied orbital centroid
and coinciding with the region of the highest electron density of the occupied orbital:
this ensures an optimal description of the correlation.

\begin{figure}[b]
\begin{center}
\includegraphics[trim=0mm 0mm 0mm 0mm,clip=true,keepaspectratio=true,width=\linewidth]{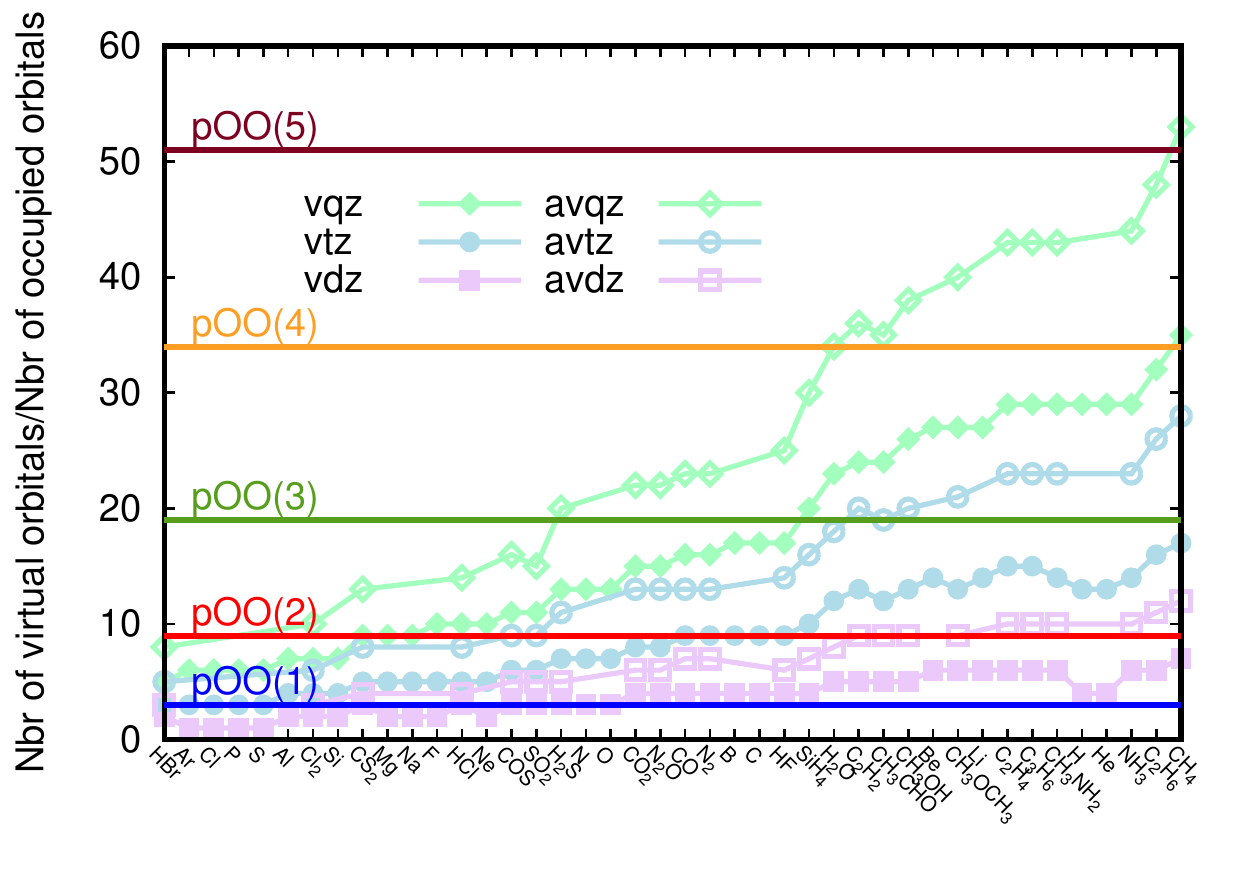}
\caption{Number of virtual orbitals
in terms of multiples of the number of occupied orbitals
for the atoms and molecules considered in this paper.
In filled symbols are the cc-pVXZ basis sets, denoted ``vdz'', ``vtz'' and ``vqz'',
and in shallow symbols are the aug-cc-pVXZ, denoted ``avdz'', ``avtz'' and ``avqz''.
The sets of pOOs up to a given order each yield a constant number of virtual orbitals
in terms of multiples of the number of occupied orbitals, shown by the horizontal lines.
}
\label{fig:virtual}
\end{center}
\end{figure}

\subsection*{Higher order pOOs}

Higher order pOOs that have further nodes
can be obtained in a similar way, using higher order polynomials.
A pOO of general order $n$ hence reads: 
\begin{align}
    \ket{i_{\alpha\dots\eta}}
    =\hat{P}\,(\hat{r}_\alpha-D^i_\alpha)\dots(\hat{r}_\eta-D^i_\eta)\ket{i}
\end{align}
It can be shown (see Appendix) that pOOs can be written as a function of pOOs of lower order:
\begin{align}
\label{eq:asfct}
    \ket{i_{\alpha\dots\eta}}
    =\hat{P}\,\hat{r}_\alpha\dots\hat{r}_\eta\ket{i}
    -\sum_{k=1}^{n-1} 
    \underbrace{\vphantom{\ket{K}}D^i \dots D^i}_{n-k}
    \underbrace{\ket{K}}_{\substack{\text{pOO of}\\\text{order }k}}
\end{align}
In this equation, every term of order $k$ in the sum is actually replicated ${n \choose k}$ times,
each time assigning a different set of $k$ indices to the pOO $\ket{K}$
and the remaining indices to the vector component $\b{D}^i$.
This explains why the terms in the sum are not written with their indices,
and might be made clearer with an example, found in the Appendix.
As a result of Eq.~(\ref{eq:asfct}),
the overlaps involving any order of pOOs are functions of lower order overlaps:
\begin{align}
 \braket{N}{M}
 &= \expect{i}{\underbrace{\hat{r}\dots\hat{r}}_{n}
 \hat{P}\,\underbrace{\hat{r}\dots\hat{r}}_{m}}{j}
\nonumber\\&
 -\underset{k+l\neq n+m}{\sum_{k=1}^{n}\sum_{l=1}^{m}}
 \underbrace{D^i\dots D^i}_{n-k} \braket{K}{L}
 \underbrace{D^j\dots D^j}_{m-l} 
\end{align}
where again the terms of the sum are replicated ${n \choose k}{m \choose l}$ times
with a different set of $k$ indices assigned to the pOO $\bra{K}$,
a different set of $l$ indices assigned to the pOO $\ket{L}$,
and the remaining indices to the vectors $\b{D}^i$ and $\b{D}^j$.
The pOOs of a given order $n$ and all equations involving those pOOs hence only need
ingredients (overlap, etc\dots) already constructed for lower order pOOs
plus at most the matrix elements of the multipole operator of order $2n$.

\subsection*{Size of the virtual orbital space}

\begin{table}[b]
\newcolumntype{C}[1]{>{\centering\arraybackslash}p{#1}}
\caption{\label{tab:nbrpOO}
Number of virtual orbitals generated at and up to a given order of harmonics used.}
\begin{tabularx}{.80\linewidth}{C{1cm}|C{14mm}C{20mm}C{20mm}}
  \hline\hline
  order of the pOOs & figurate number & number of pOOs generated & cumulant number of pOOs\\
  \hline
  \\[-3mm]
  1 & $\displaystyle{3 \choose 1}$ & $\phantom{1}3\times N_\text{LMO}$ & $\phantom{1}3\times N_\text{LMO}$\\[5mm]
  2 & $\displaystyle{4 \choose 2}$ & $\phantom{1}6\times N_\text{LMO}$ & $\phantom{1}9\times N_\text{LMO}$\\[5mm]
  3 & $\displaystyle{5 \choose 3}$ & $          10\times N_\text{LMO}$ & $          19\times N_\text{LMO}$\\[5mm]
  4 & $\displaystyle{6 \choose 4}$ & $          15\times N_\text{LMO}$ & $          34\times N_\text{LMO}$\\[5mm]
  5 & $\displaystyle{7 \choose 5}$ & $          21\times N_\text{LMO}$ & $          51\times N_\text{LMO}$\\[4mm]
  \hline\hline
\end{tabularx}
\end{table}

\begin{figure*}[t]
\begin{center}
   \setlength\fboxrule{1pt}
   \setlength\fboxsep{0pt}
\begin{minipage}{.2\linewidth}
   \includegraphics[width=\textwidth]{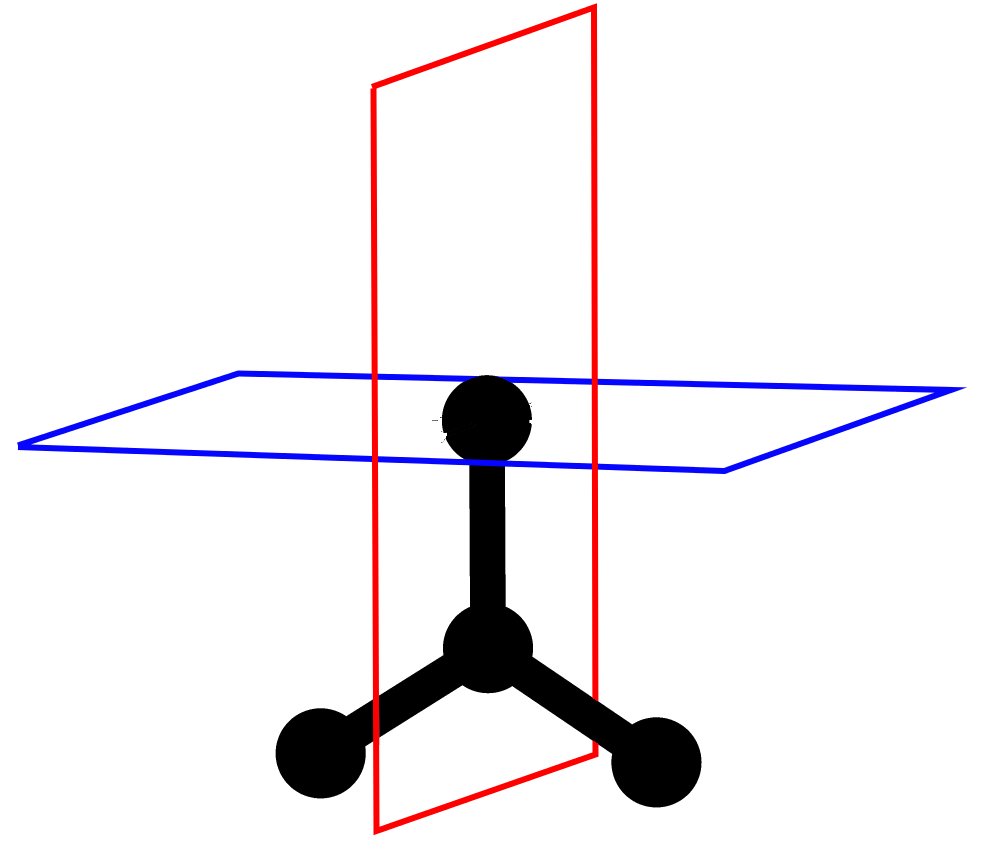}
\end{minipage}\begin{minipage}{.8\linewidth}
   \subfigure{\fbox{\includegraphics[width=0.22\textwidth]{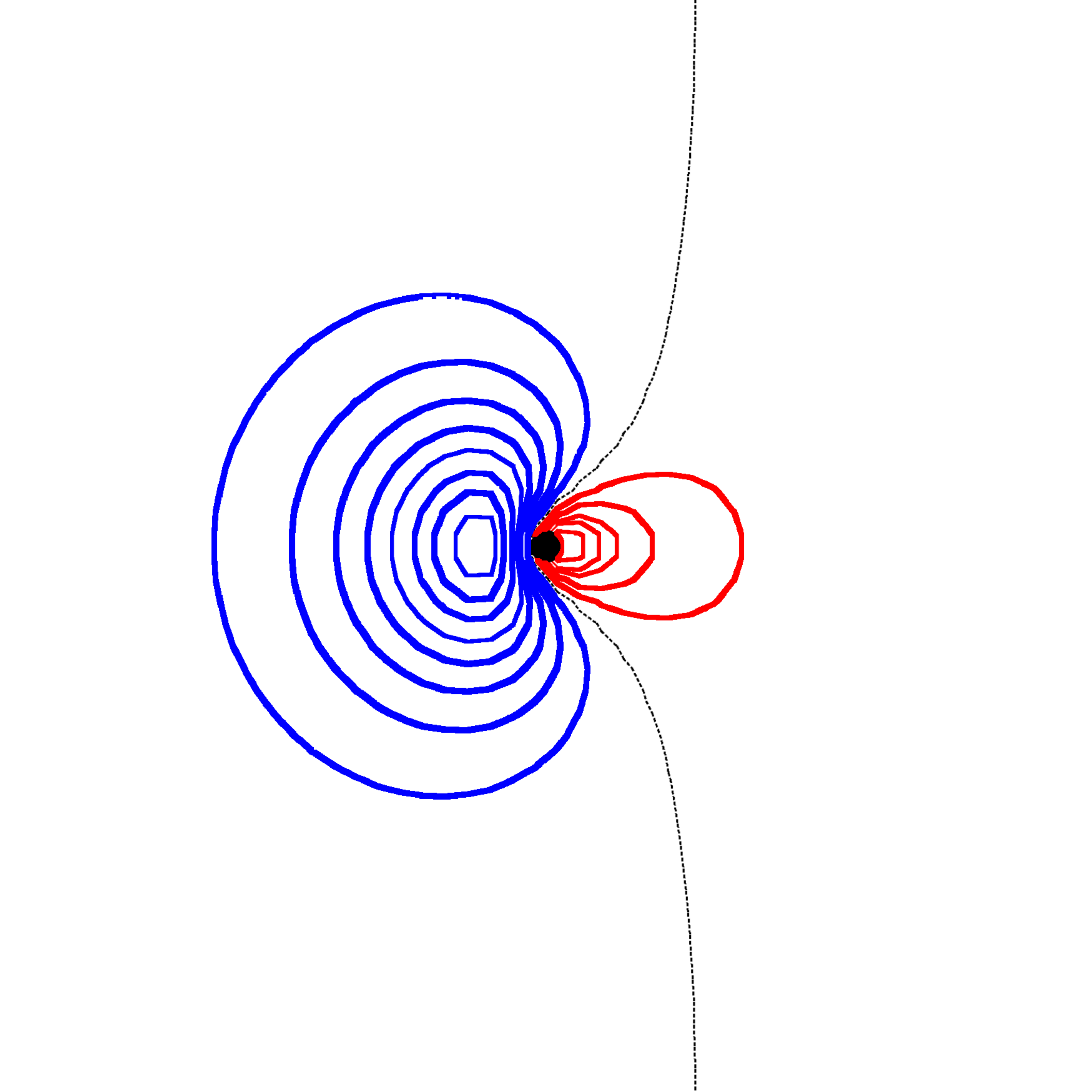}}}
   \subfigure{\fbox{\includegraphics[width=0.22\textwidth]{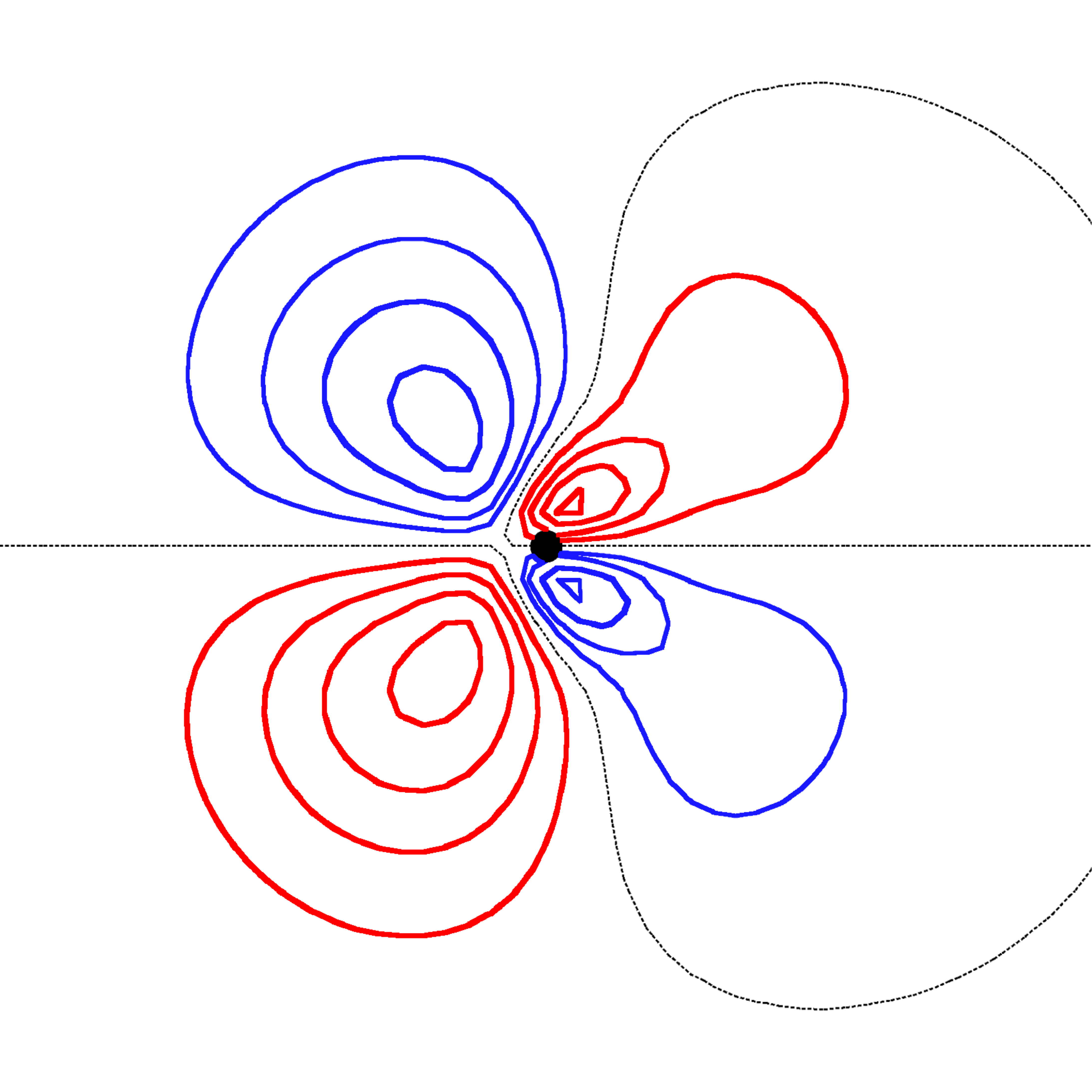}}}
   \subfigure{\fbox{\includegraphics[width=0.22\textwidth]{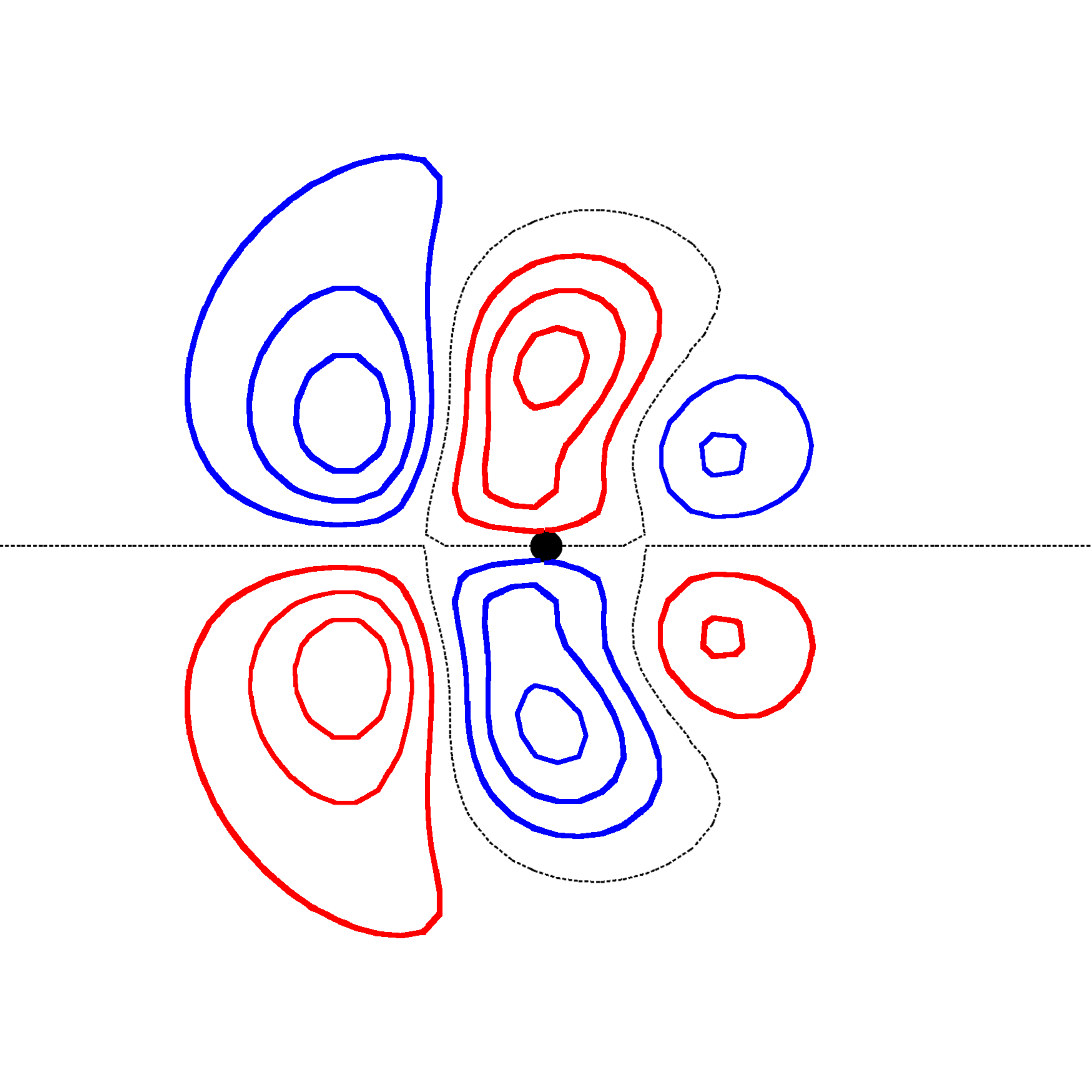}}}
   \subfigure{\fbox{\includegraphics[width=0.22\textwidth]{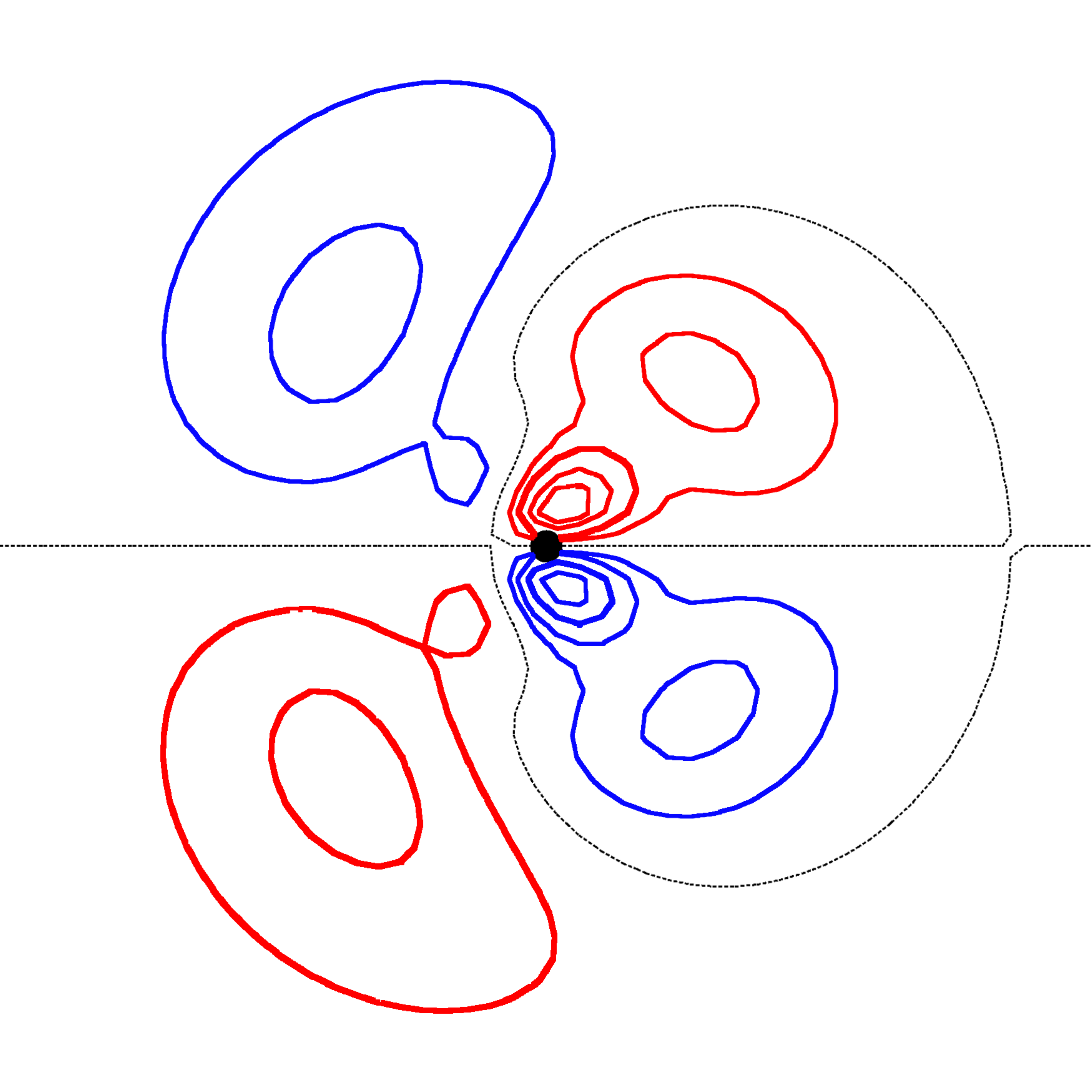}}}

   \vspace*{-.3em}

   \subfigure{\fbox{\includegraphics[width=0.22\textwidth]{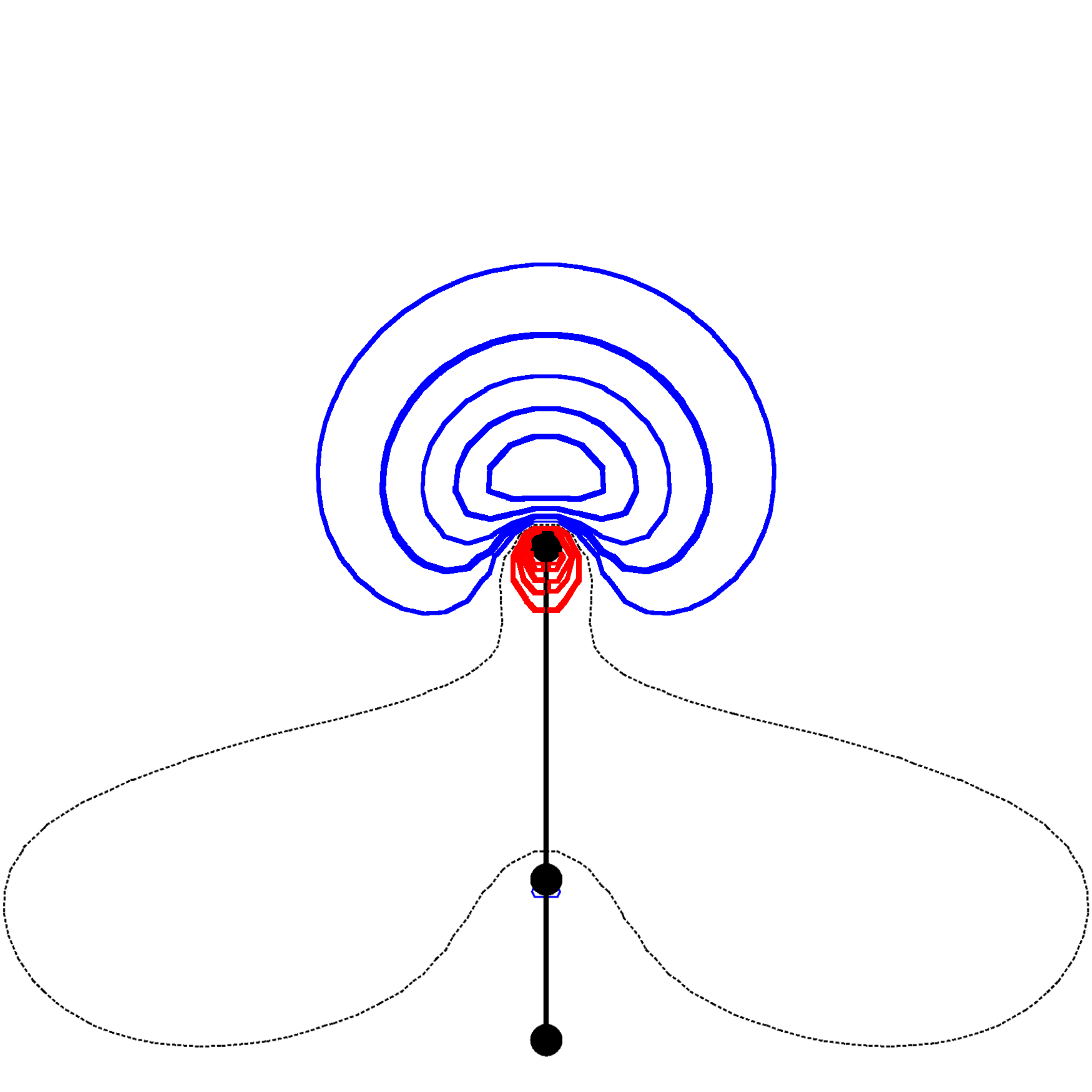}}}
   \subfigure{\fbox{\includegraphics[width=0.22\textwidth]{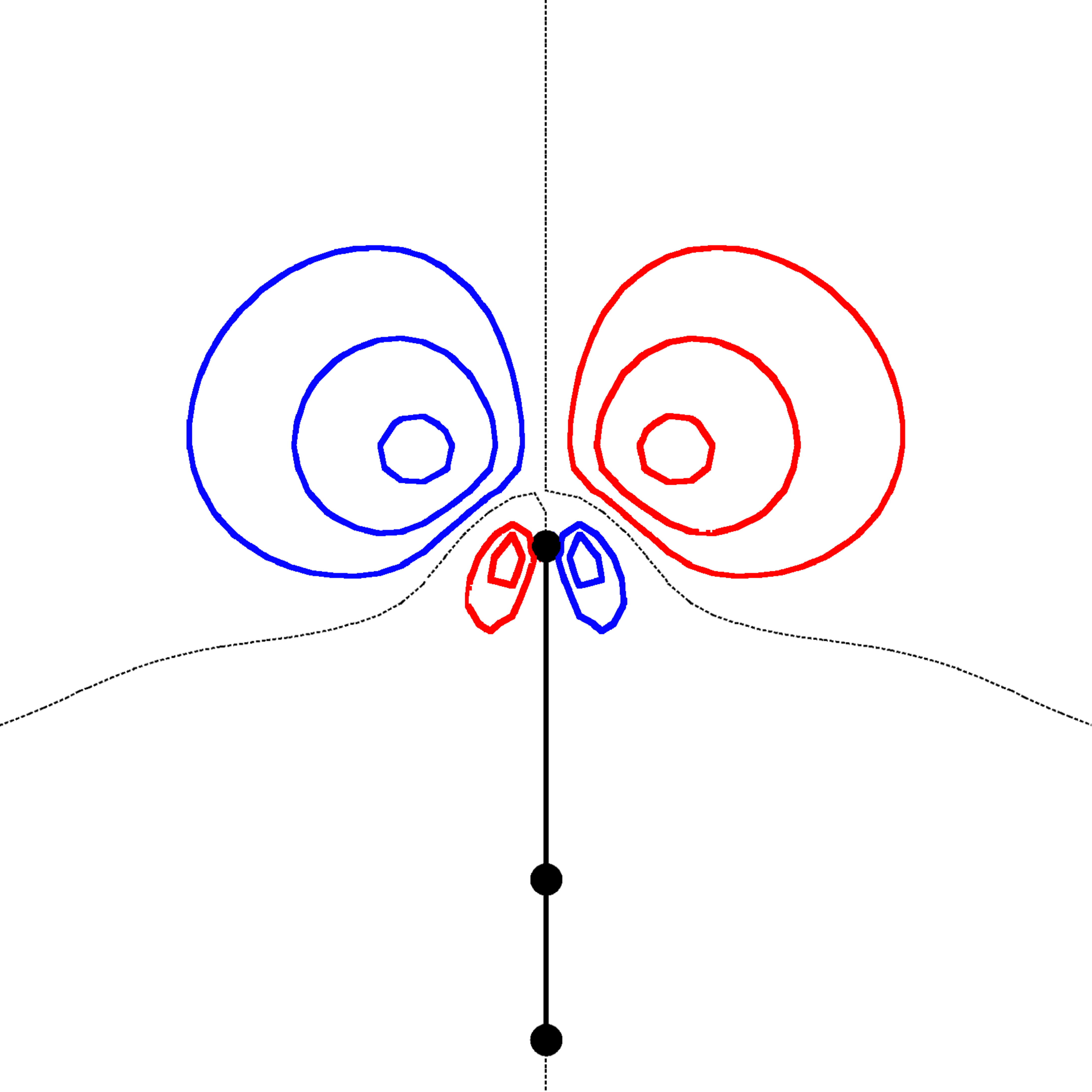}}}
   \subfigure{\fbox{\includegraphics[width=0.22\textwidth]{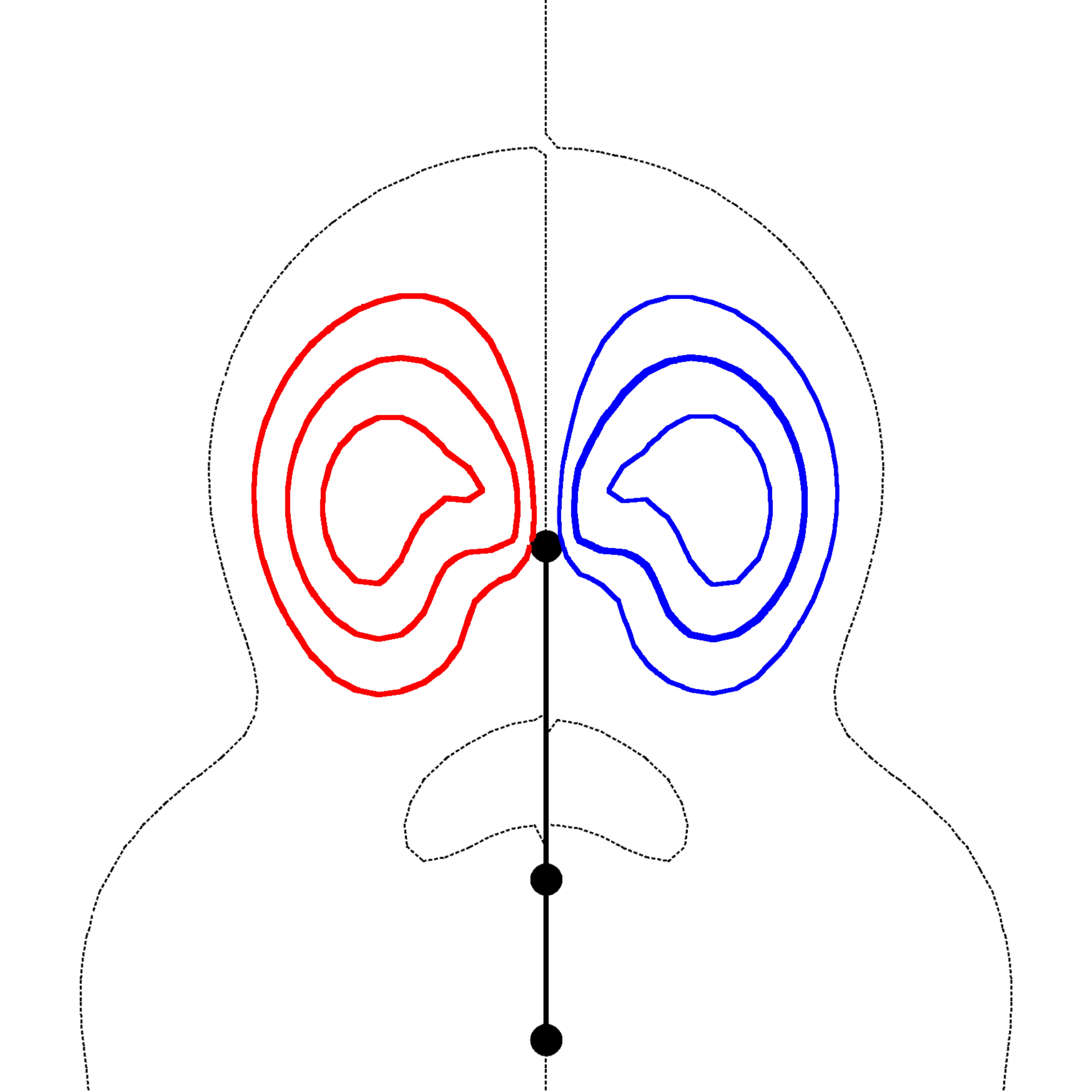}}}
   \subfigure{\fbox{\includegraphics[width=0.22\textwidth]{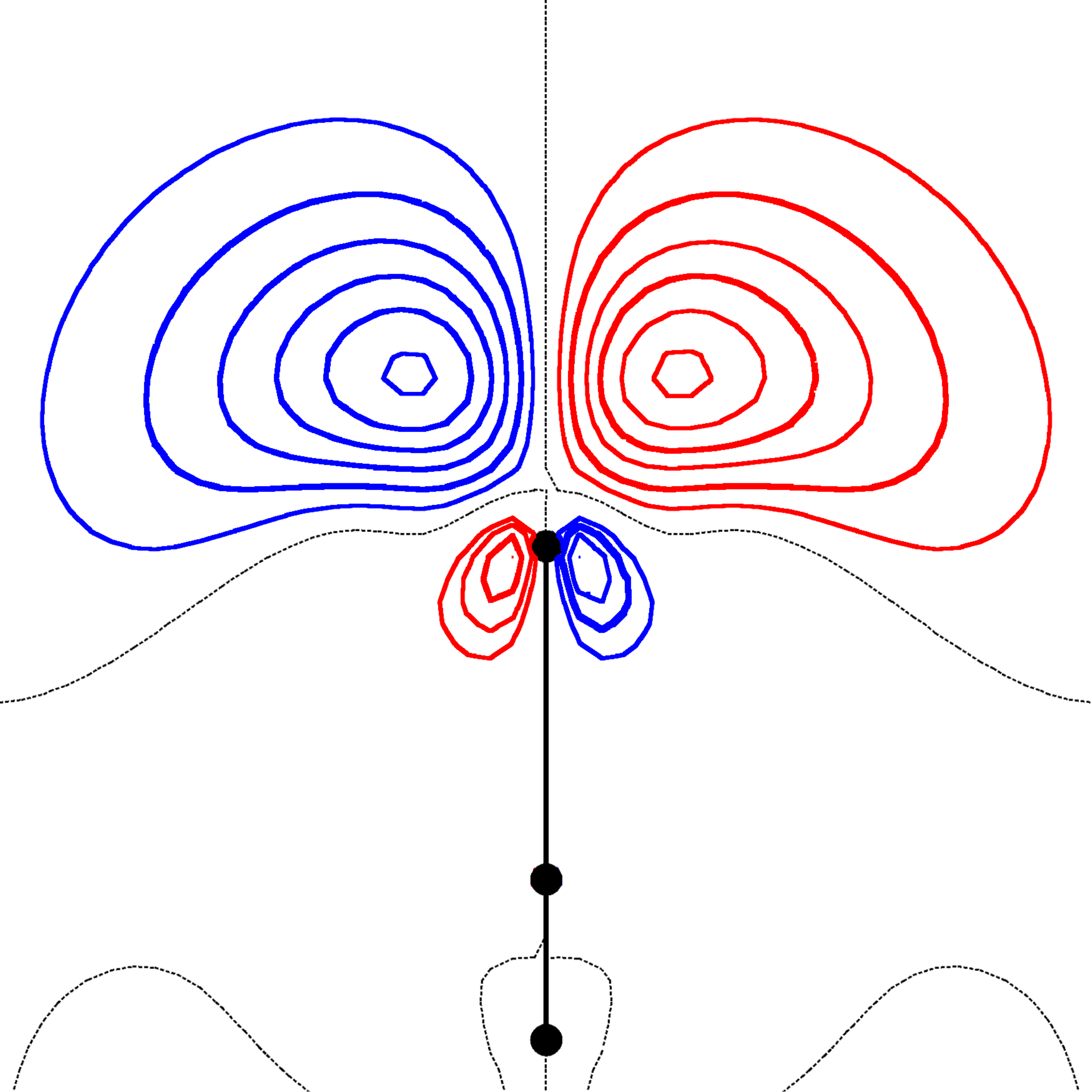}}}

\end{minipage}
\caption{\label{pOOs}
Examples of pOOs generated from the oxygen lone pair of the formaldehyde (left panels)
by applying first-, second- and third order harmonics (second, third and last panels).
The first row shows contour taken on the blue plane in the sketch molecule on the left,
the second row shows contour taken on the red plane.
}
\end{center}
\end{figure*}

The number of first order pOOs generated is $3.N_\text{LMO}$
(the $x$, $y$ and $z$ pOOs constructed for each LMO).
In terms of multiples of the number of LMOs,
the number of pOOs generated at each order $n$ of spherical harmonics used is given by a figurate number
(linear number for first order,
triangular number for second order,
tetrahedral number for third order, etc\dots),
which simply corresponds to the number of unique $n$-tuplets that can be made with $x,y,z$
\begin{align}
  P(n)=\frac{1}{2}(n+1)(n+2)={n+2 \choose n}
\end{align}
This is summed-up in Table~\ref{tab:nbrpOO},
which also shows the cummulated number of pOOs generated \textit{up to} order $n$ for the first 5 orders.
As comparison, Figure~\ref{fig:virtual} shows the number of virtual orbital generated by
the widely used Dunning basis sets cc-pVDZ, cc-pVTZ, cc-pVQZ, aug-cc-pVDZ, aug-cc-pVTZ, and aug-cc-pVQZ
for the atoms and molecules on which the correlation energy are calculated in this paper
(those are taken from the set of the small systems compiled
by Tkatchenko and Scheffler\cite{Tkatchenko:09,Toulouse:13}).
This shows that for example the set of pOOs up to third order yields
a size of virtual space greater than the one generated
by the cc-pVTZ and on par with the one generated by aug-cc-pVTZ.

\section*{Local RPA equations}

The random phase approximation can be formulated
in a number of different ways\cite{Angyan:11,Toulouse:11,urpa,diel,frac}, but 
in a local formulation of the ``ring coupled-cluster double'' (rCCD),
the RPA correlation energy can be given by a sum over (localized) pair contributions:
\begin{align}
E_c=\frac{1}{2}\text{Tr}\left(\b{K}^{ij}\b{T}^{ji}\right)
\end{align}
where the amplitudes $\b{T}$ are found by solving the Riccati equations $\b{R}^{ij}=\b{0}$, where:
\begin{align}
\b{R}^{ij}
&=\b{B}^{ij}
 +\b{A}^{im}\b{T}^{mj}\b{S}
 +\b{S}\b{T}^{im}\b{A}^{mj}
\nonumber\\&\quad
 +\b{S}\b{T}^{im}\b{B}^{mn}\b{T}^{nj}\b{S}
\end{align}
where the matrices are written in the LMOs/pOOs basis:
\begin{align}
A^{ij}_{k_\alpha l_\beta} &= \braket{ij}{k_\alpha l_\beta} - \braket{k_\alpha i}{l_\beta j}
                            +\epsilon^{ij}_{k_\alpha l_\beta}
\nonumber\\
B^{ij}_{k_\alpha l_\beta} &= \braket{ij}{k_\alpha l_\beta} - \braket{ij}{l_\beta k_\alpha}
\nonumber\\
\epsilon^{ij}_{k_\alpha l_\beta} &= \delta_{ij}f_{k_\alpha l_\beta}-S_{k_\alpha l_\beta} f_{ij}
\end{align}
where $\b{f}$ is the Fock matrix in LMOs/pOOs basis and
the two-body integrals between LMOs and pOOs are in physicist's notation.
The Riccati equations are solved iteratively in a pseudo-canonical basis;
see our original paper\cite{us} for more detailed derivations and a detailed workflow in the appendix.

\begin{figure*}[!hbtp]
\begin{center}
\includegraphics[width=.45\linewidth]{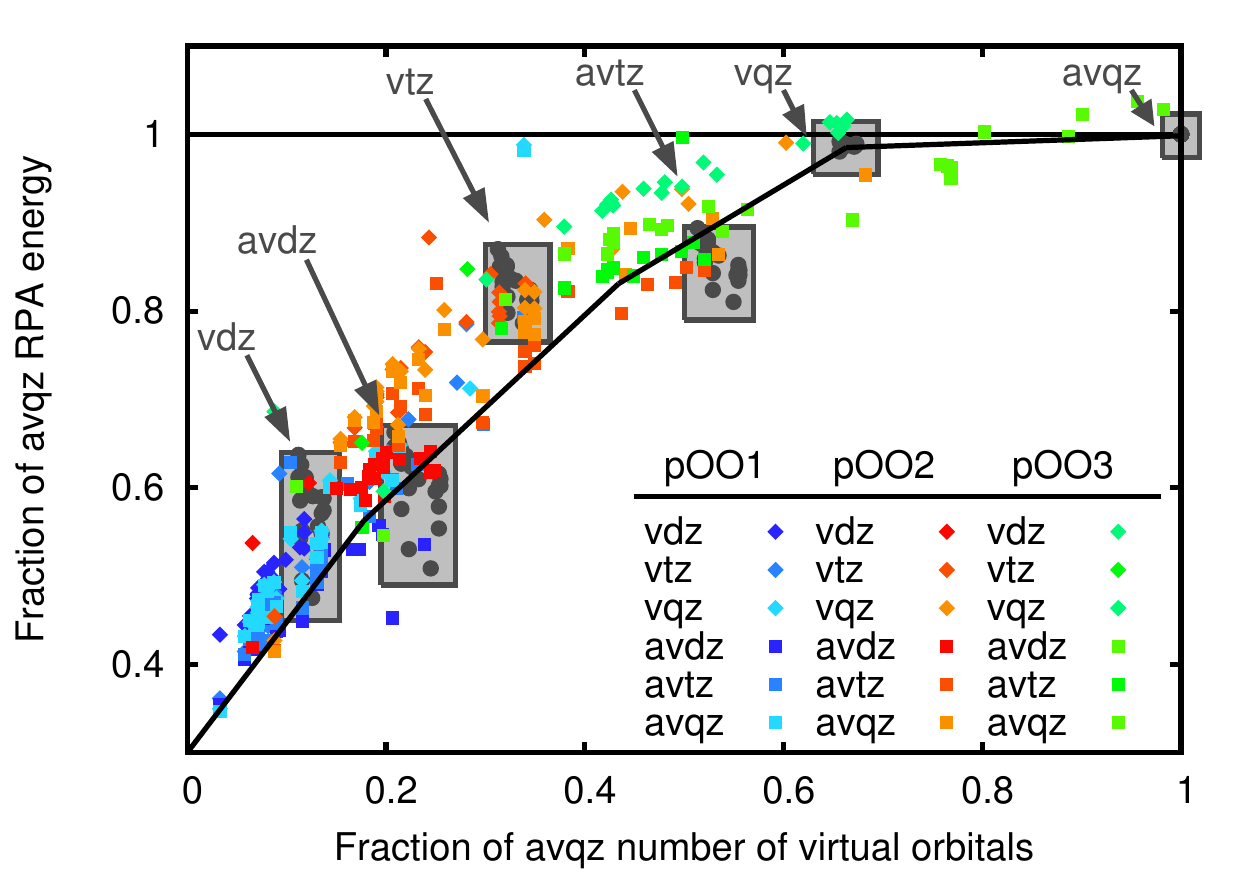}
\includegraphics[width=.45\linewidth]{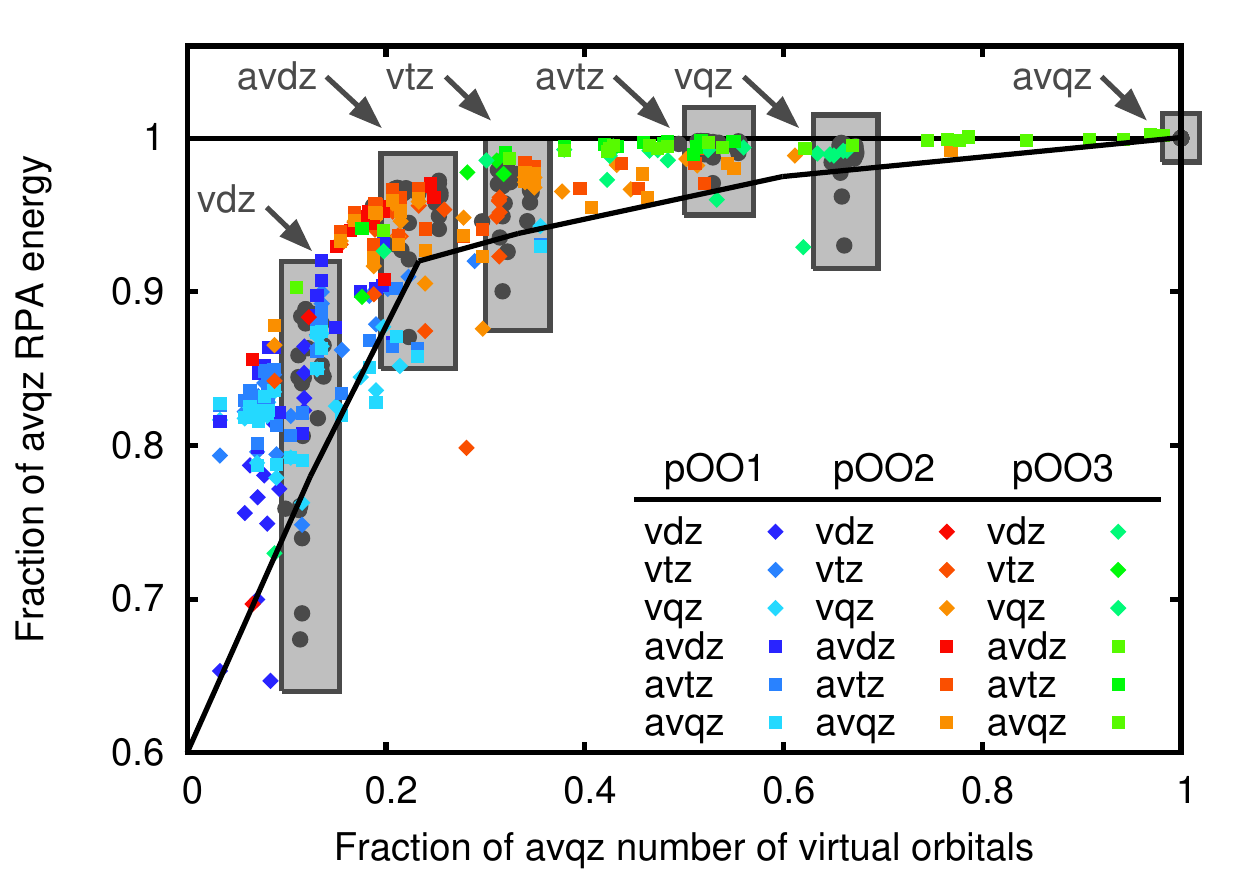}
\caption{
Results of calculations of the RPA correlation energies (left: full-range, right: long-range)
of a small set of molecules.
Each dot represents a result for one of those molecule, obtained with several different basis.
The gray boxes gather the gray dots corresponding to reference results
that are obtained with the unmodified cc-pVDZ, aug-cc-pVDZ,
cc-pVTZ, aug-cc-pVTZ, cc-pVQZ and aug-cc-pVQZ basis sets (denoted as vdz, avdz, etc\dots).
The color dots are the pOO results, with pOOs of first (second and third) order being in blue (red and green).
Those pOOs can be obtained in the context of the different reference Dunning basis,
hence the nuances of blue, red and green.
The black line in both plots roughly connects the centers of the gray boxes.
\label{fig:Ec}
}
\end{center}
\end{figure*}

\section*{Correlation energies}

The performance of the pOOs is investigated for the calculation of
the RPA correlation energies of a set of small molecules\cite{Tkatchenko:09}
already used in our original paper\cite{us}.
Both post-HF calculations
and range-separated hybrid calculations using the RSHPBE functional with a parameter $\mu=0.5$ are performed.
The occupied orbitals obtained are localized, and the pOOs are generated before
calculating either the full-range or the long-range RPA correlation energies.

All calculations were performed with a development version of
the quantum chemical program package Molpro\cite{Molproshort:10},
where the construction of the pOOs up to third order has been implemented for the purpose of this paper,
together with the corresponding overlaps.

Results of calculations using the cc-pVDZ, cc-pVTZ, cc-pVQZ
as well as the aug-cc-pVDZ, aug-cc-pVTZ and aug-cc-pVQZ basis sets are shown.
In practice, because of linear dependencies built in the pOOs (see briefly the workflow in the Appendix),
the RPA equations are carried with an effective number of independant pseudo-canonical basis functions\cite{us}
which can be lower than the multiple of the number of occupied orbitals mentioned in Table~\ref{tab:nbrpOO} and in Figure~\ref{fig:virtual}.
The pertinent quantities to look at are hence
the RPA correlation energies and the number of pseudo-canonical orbitals involved in the calculations.
Figure~\ref{fig:Ec} shows this in terms of the fractions of
the aug-cc-pVQZ RPA correlation energies recovered
as a function of the fractions of the aug-cc-pVQZ number of virtual orbitals involved in the calculations.
On the left are shown post-HF results and on the right long-range RPA correlation energy results.

The RPA correlation energies obtained in each basis sets
without the use of localized orbitals are shown in gray dots and gathered for clarity within gray boxes.
The rest of the data
(the RPA correlation energies obtained when using LMOs and pOOs) is shown in colored symbols.
The results from pOOs obtained within a (aug-)cc-pVXZ basis are shown in diamonds (squares).
The pOOs of first order are shown in shades of blue,
of second order in shades of red and of third order in shades of green.

A first interesting result is that the performances of a given order of pOOs is not significantly changed
by the basis used to construct them: pOOs will perform a certain way
independently of the core basis set used
(this is even more striking in the case of the long-range results,
right of Figure~\ref{fig:Ec}).
Furthermore, a positive finding is that the calculations involving pOOs recover
a good portion of the targeted RPA correlation energies (full- or long-range)
for the size of the virtual space involved.
As a marker, the black line in the figure is drawn roughly between the center of the gray boxes
holding the non-localized RPA energies, 
and shows the fraction of the aug-cc-pVQZ correlation energy recovered by smaller basis sets.
The pOO results lie virtually systematically above that line,
meaning the pOOs recover a larger portion of the targeted RPA energies
than the basis set having similar numbers of virtual orbitals.
Finally, as expected, the fraction of long-range RPA correlation energy recovered is significantly higher
than the recovered portion of full-range correlation energy.
This is a nice result, as the framework of pOOs was largely thought as a way to build approximations
to model the London dispersion forces, as can be seen in our original paper\cite{us}.

\section*{Conclusion}

It was shown that localized virtual orbitals can be constructed easily and systematically
by multiplying localized occupied orbitals by harmonic functions of higher than first order.
The size of the virtual space obtained is
comparable to the size of that generated when using the Dunning basis sets,
and the RPA correlation energies calculated
with the pOO localized virtual orbitals are close to the energies without local approximations.
Because of an interesting iterative feature emerging
in the construction of the projected oscillator orbitals,
it is easy to generate higher order pOO from generated lower order pOOs.
A great number of additional approximation, developed in our original paper\cite{us} in the context
of first order projected oscillator orbitals,
such that multipolar expansion to approximate the electron repulsion, spherical approximation, etc.
could be applied in the future to the higher order pOOs derived here.

\section*{Acknowledgements}

This research was conducted while in the University of Colorado at Boulder
and hence was supported through the startup package of Sandeep Sharma.
The author would like to thank the late János G. Ángyán
for the initial insightful discussions on this subject and on many others.
János was a great mentor to me and he is missed very much.

\appendix
\section*{Appendix}

\subsection*{Relations between pOOs}

The natural expression for a pOO of order $n$ is given by the binomial theorem:
\begin{align}
\label{eq:binom}
\ket{i_{\alpha\dots\eta}}
&=\hat{P}\,
  (\hat{r}_\alpha-D^i_\alpha)\dots(\hat{r}_\eta-D^i_\eta)\ket{i}
=
  \sum_{k=1}^{n} A^n_k
\end{align}
where $A^n_k$ is defined for convenience:
\begin{align}
A^n_k
 =(-1)^{n-k}\underbrace{(D^i\dots D^i)}_{n-k}
  \hat{P}\,
  \underbrace{\vphantom{()}\hat{r}\dots\hat{r}}_{k}\ket{i}
\end{align}
and actually contains ${n \choose k}$ repetitions
with different assignations of the indices $\alpha\dots\eta$.
For each of those repetitions, a different set of $k$ indices among $\alpha\dots\eta$
is assigned to the $k$ position operators
and the remaining indices are given to the vector components $D^i$.
For example, consider a pOO $\ket{i_{\alpha\beta\gamma}}$ of order 3.
The term $A^3_1$ will be:

\begin{align}
A^3_1{}^{\phantom{\prime\prime}}
&=(D^i_\alpha D^i_\beta)  \hat{P}\,\hat{r}_\gamma\ket{i}
\nonumber\\
A^3_1{}^{\prime\phantom{\prime}}
&=(D^i_\alpha D^i_\gamma) \hat{P}\,\hat{r}_\beta \ket{i}
\nonumber\\
A^3_1{}^{\prime\prime}
&=(D^i_\beta  D^i_\gamma) \hat{P}\,\hat{r}_\alpha\ket{i}
\nonumber
\end{align}
\textit{i.e.} it contains ${3 \choose 1}=3$ repetitions with different assignations
of the indices $\alpha$, $\beta$ and $\gamma$.

We seek to prove Eq.~(\ref{eq:asfct}), which amounts with the current notations to:
\begin{align}
\label{eq:seek}
    \ket{i_{\alpha\dots\eta}}
    =A^n_n
    -\sum_{k=1}^{n-1} 
    \underbrace{\vphantom{\ket{nk}}D^i \dots D^i}_{n-k}
    \underbrace{\ket{K}}_{\substack{\text{pOO of}\\\text{order }k}}
\end{align}
where the same repetitions
with different assignations of the indices $\alpha\dots\eta$
are found now for each terms of the sum.
We begin by inserting in Eq.~(\ref{eq:seek})
the binomial expression of Eq.~(\ref{eq:binom}) for the pOO $\ket{K}$:
\begin{align}
\ket{i_{\alpha\dots\eta}}
&=A^n_n
  -\sum_{k=1}^{n-1}
  \underbrace{(D^i\dots D^i)}_{n-k}
  \sum_{m=1}^k A_m^k
\end{align}
where we can rewrite the term in the sums using:
\begin{align}
&\underbrace{(D^i\dots D^i)}_{n-k}  A_m^k
\nonumber\\&\quad
    = (-1)^{k-m}
    \underbrace{(D^i\dots D^i)}_{n-k} \underbrace{(D^i\dots D^i)}_{k-m}
    \hat{P}\,
    \underbrace{\vphantom{(}\hat{r}\dots\hat{r}}_{m}\ket{i}
  \label{eq:ways}
\\&\quad
    =(-1)^{k+n}A^n_m
\nonumber
\end{align}
This yields:
\begin{align}
\ket{i_{\alpha\dots\eta}}
&=A^n_n
  +\sum_{k=1}^{n-1} \sum_{m=1}^k (-1)^{k+(n-1)}
  A^n_m
\nonumber\\
&=A^n_n
  +\sum_{m=1}^{n-1} \left[\sum_{k=m}^{n-1} (-1)^{k+(n-1)}\right]
  A^n_m
\label{eq:Am}
\end{align}
where in the last line the double summation was simply rearranged
(think of summations over lines versus over columns of a matrix whose columns are composed of $A^n_m$).

In terms of assignations of the indices $\alpha\dots\eta$, in Eq.~(\ref{eq:ways})
there is ${n \choose k}$ ways to assign indices to the first string of $n-k$ ``$D^i$''
and ${k \choose m}$ ways to assign indices to the second string of $k-m$ ``$D^i$''.
This hence becomes a combinatorial problem,
and recovering Eq.~(\ref{eq:binom}) from Eq.~(\ref{eq:Am}) boils down to proving that:
\begin{align}
 \sum_{k=m}^{n-1} (-1)^{k+(n-1)} {n\choose k}{k\choose m}={n\choose m}
\end{align}
which should be the number of repeated occurences of
the $A^n_m$ term with different indices in Eq.~(\ref{eq:Am}).
This amounts to prove that:
\begin{align}
    \sum_{k=m}^{n-1} (-1)^{k+(n-1)} \frac{{n\choose k}{k\choose m}}{{n\choose m}}=1
\end{align}
where, manipulating the fraction, the expression to study is:
\begin{align}
&\sum_{k=m}^{n-1} (-1)^{k+(n-1)} {n-m\choose k-m}
\nonumber\\
&=\sum_{k=m}^{n}   (-1)^{k+(n-1)} {n-m\choose k-m} - (-1)^{-1} {n-m\choose n-m}
\nonumber\\
&=\left\{\sum_{k^\prime=0}^{n-m} (-1)^{k^\prime} {n-m\choose k^\prime}\right\} (-1)^{n-m+1} + 1
\nonumber\\
&=1
\end{align}
where the first step consists simply in adding and substracting the $n$-th element of the sum,
the second step is a change of the dummy index of the sum
and the final step is a realization that the expression in curvy brackets is
the binomial theorem for $(1+(-1))^{n-m}=0$.
This proves Eq.~(\ref{eq:asfct}).

\subsection*{Workflow}

Although in our original paper\cite{us} relations to construct all ingredients for the pOOs
using solely data from the occupied space are presented (this is the purpose of the projected schemes),
for the clarity of the proof-of-concept calculations in this paper,
the $N_\text{pOO}\times N_\text{vir}$ matrix $\b{V}$ to transform virtual orbitals to pOOs is introduced:
\begin{align}
    \ket{i_{\alpha\dots\eta}}=\sum_a^{N_\text{vir}} V_{i_{\alpha\dots\eta} a} \ket{a}
\end{align}
We assume here that the pOOs can be exanded in the virtual basis,
\textit{i.e.} that the pOO space is a subspace of the virtual space.
This should be more rigourously checked in later work.
For example, the $\b{V}$ matrix corresponding to the first order pOO reads:
\begin{align}
V^{(1)}_{i_\alpha a}=\expect{i}{\hat{r}_\alpha}{a}
\end{align}
and, given Eq.~(\ref{eq:asfct}), repeated in the Appendix as Eq.~(\ref{eq:seek}),
the $\b{V}$ matrix corresponding to a $N$-th order pOO simply reads:
\begin{align}
    V^{(N)}_{i_{\alpha\dots\eta} a}=\expect{i}{\hat{r}_\alpha\dots\hat{r}_\eta}{a}
    -\sum_{k=1}^{n-1} 
    \underbrace{\vphantom{V_a}D^i\dots D^i}_{n-k} 
    \underbrace{V^{(K)}_{\dots a}}_{\substack{\text{matrix for}\\\text{order $k$}\\\text{pOO}}}
\end{align}
The overlap between pOOs is $\b{S}=\b{V}\b{V}^\dagger$ and the pseudo-canonical basis
is obtained by solving the generalized eigenvalue equation
\begin{align}
\b{f}\b{X}=\b{S}\b{X}\B{\epsilon}
\end{align}
Because of linear dependencies built in the construction of the pOOs,
the overlap matrix $\b{S}$ is in general only positive semidefinite instead of positive definite.
Hence the pseudo-canonical basis of effective size $N_\text{eff}\le N_\text{pOO}$ can alternatively
by found with the $N_\text{pOO}\times N_\text{eff}$ matrix $\b{X}$ that
similtaneously diagonalizes $\b{f}$ and $\b{S}$.

From these relations, one can then construct from the $N_\text{AO}\times N_\text{vir}$
tranformation matrix $\b{C}$ the following sequential transformation matrices:
\begin{align}
  \underbrace{\b{C}}_{\text{AO}\rightarrow \text{vir}}
  \rightarrow
  \underbrace{\b{C}\b{V}^\dagger}_{\text{AO}\rightarrow \text{pOO}}
  \rightarrow
  \underbrace{\b{C}\b{V}^\dagger\b{X}}_{\text{AO}\rightarrow \text{pseudo-cano}}
\end{align}
to easily either try out the pOOs, draw them, or calculate energies.

\end{document}